\documentclass[prc,twocolumn]{revtex4-1}
\usepackage[latin1]{inputenc}
\usepackage{bm}
\usepackage{epsfig}
\usepackage{amsthm}
\usepackage{dcolumn}
\usepackage{graphicx}
\usepackage{amsfonts}
\usepackage{float}
\usepackage{amsmath,amssymb}

\def\qb{\bar{q}}

\def\x{\bm x}
\def\y{\bm y}
\def\k{\bm k}
\def\q{\bm q}
\def\p{\bm p}
\def\P{\bm P}
\def\bsigma{\bm\sigma}
\def\balpha{\bm\alpha}
\def\bnabla{\bm\nabla}

\def\beq{\begin{equation}}
\def\eeq{\end{equation}}
\def\bea{\begin{eqnarray}}
\def\eea{\end{eqnarray}}
\def\nn{\nonumber}

\begin{document}
\title{$\bar D N$ interaction in a color-confining chiral quark model}
\author{C.E. Fontoura, G. Krein and V.E. Vizcarra}
\affiliation{Instituto de F\'{\i}sica Te\'orica, Universidade Estadual Paulista\\
Rua Dr. Bento Teobaldo Ferraz, 271 - Bloco II, 01140-070 S\~ao Paulo, SP, Brazil}
\begin{abstract}
We investigate the low-energy elastic $\bar DN$ interaction using a quark
model that confines color and realizes dynamical chiral symmetry
breaking. The model is defined by a microscopic Hamiltonian inspired
in the QCD Hamiltonian in Coulomb gauge. Constituent quark masses
are obtained by solving a gap equation and baryon and meson
bound-state wave functions are obtained using a variational method. We
derive a low energy meson-nucleon potential from a quark-interchange
mechanism whose ingredients are the quark-quark and quark-antiquark
interactions and baryon and meson wave functions, all derived from the same
microscopic Hamiltonian. The model is supplemented with
($\sigma$,~$\rho$,~$\omega$,~$a_0$) single-meson exchanges to describe
the long-range part of the interaction. Cross-sections and phase shifts
are obtained by iterating the quark-interchange plus meson-exchange potentials
in a Lippmann-Schwinger equation. Once model parameters in meson exchange
potential are fixed to describe the low-energy experimental phase shifts of
the $K^+N$ and $K^0 N$ reactions, predictions for $\bar D^0 N$ and
$D^- N$ reactions are obtained without introducing new parameters.
\end{abstract}

\maketitle

\section{Introduction}
\label{sec:intro}

The interaction of heavy-flavored hadrons with ordinary hadrons like
nucleons and light mesons is focus of great contemporary interest in
different contexts. One focus of interest is in experiments of
relativistic heavy ion collisions. In~heavy ion experiments, charm
and bottom quarks are produced in the initial stages of the
collision by hard scattering processes. Since they are much heavier
than the light partons making up the bulk of the matter produced in
the collision, it is likely that they will not equilibrate with the
surrounding matter and therefore they might be ideal probes of
properties of the expanding medium -- for a recent review on the
subject and an extensive list of references see
Ref.~\cite{Rapp:2008tf}. However, the heavy quarks will eventually
hadronize and information on the medium will be carried out of the
system by heavy-flavored hadrons. In their way out of the system,
the heavy-flavored hadrons will interact with the more abundant
light-flavored hadrons and a good unders\-tan\-ding of the interaction
is crucial for a reliable interpretation of experimental data.
Another focus of in\-te\-rest is an exciting physics program that
will be carried out with the $12$~GeV upgrade of the CEBAF
accelerator at the Jefferson Lab in the USA and the construction of
the FAIR facility in Germany. At the Jefferson Lab charmed hadrons
will be produced by scattering electrons off nuclei and at FAIR they
will be produced by the annihilation of antiprotons on nuclei. One
particularly exciting perspective is the possibility of creating new
e\-xo\-tic nuclear bound states by nuclei capturing charmonia states
like $J/\Psi$ and
$\eta_c$~\cite{{Brodsky:1989jd},{Ko:2000jx},{Krein:2010vp},{Tsushima:2011kh}},
or heavy-light $D$ and $D^*$
mesons~\cite{{Tsushima:1998ru},{GarciaRecio:2010vt},
{GarciaRecio:2011xt}}. In addition, the quest for experimental
signals of chiral symmetry restoration in matter is a subject of
immense current interest and open-charm $D$ and $D^*$ mesons are
expected to play an important role in this respect -- for a recent
review, see Ref.~\cite{hatsuda-rmp}. In $D$ mesons, properties of
the light constituent quarks determined by dynamical chiral symmetry
breaking (D$\chi$SB), like masses and magnetic moments, are
sensitive to the surrounding environment and changes in such 
properties will impact the structure and interactions of the mesons 
in medium.
Evidently, a prerequisite for reliable predictions of modifications
of hadron properties in medium is a good understanding of the
free-space interaction of these hadrons. Know\-ledge of the
interaction in free space is also essential for gui\-ding future
experiments aiming at producing exotic bound states and measuring
in-medium hadron properties.

The present paper is concerned with the low-energy interaction of 
$\bar D = (\bar D^0, D^ -)$ mesons with nu\-cleons in free space.
There is a complete lack of experimental information on this reaction,
all of what is presently known about the interaction in free space
has been apprehended from model calculations based on hadronic
Lagrangians motivated by flavor $SU(4)$
symmetry~\cite{{Mizutani:2006vq},{Lin:1999ve},{Hofmann:2005sw}},
models using hadron and quark degrees of
freedom~\cite{Haidenbauer:2007jq} and heavy-quark
symmetry~\cite{{Yasui:2009bz},{GarciaRecio:2008dp}}, and the
nonrelativistic quark model~\cite{Carames:2012bd}. Although the
first lattice QCD studies of the interaction of charmed hadrons like
$J/\Psi$ and $\eta_c$ with nucleons are starting to appear in the
literature~\cite{{Yokokawa:2006td}, {Kawanai:2010ev},{Liu:2008rza}},
the $\bar D N$ interaction does not seem to have been considered by
the lattice community. In view of this situation, the use of models
seems to be the only alternative for making the urgently needed
predictions for e.g. low-energy cross sections of reactions
involving charmed hadrons. However, as advocated in the works of
Refs.~\cite{{Haidenbauer:2007jq},{Haidenbauer:2008ff},
{Haidenbauer:2009ad},{Haidenbauer:2010ch}}, minimally reliable
predictions of unknown cross-sections need to be founded on models
constrained as much as possible by symmetry arguments, analogies
with other similar processes, and the use of different degrees of
freedom. In the specific case of the $\bar DN$ reaction,
Ref.~\cite{Haidenbauer:2007jq} extended a very successful model for
the $KN$ reaction~\cite{Hadjimichef:2002xe}, in which the long-range
part of the interaction is described within a meson-exchange
framework~\cite{Juel1,Juel2} and the short-distance part is
described by a quark-interchange mechanism from
one-gluon-exchange~(OGE) of a non\-re\-la\-ti\-vistic quark model
(NRQM)~\cite{{Barnes:1992ca},{SilvestreBrac:1995gz},
{SilvestreBrac:1997jw},{Lemaire:2001sr},{Lemaire:2002wh},{Lemaire:2003et}}.
The model of Ref.~\cite{Hadjimichef:2002xe} describes the available
low-energy experimental data for the $KN$ reaction and was used to
set limits on the width of the hypothetical $\Theta^+(1540)$
pentaquark state~\cite{Haidenbauer:2003rw}. For the $\bar D N$
reaction, the model predicts cross-sections that are on average of
the same order of magnitude but larger by a factor roughly equal to
$2$ than those of the analogous $K^+N$ and $K^0N$ reactions for 
center-of-mass kinetic energies up to $150$~MeV. Two interesting 
fin\-dings of the study of Ref.~\cite{Haidenbauer:2007jq} are 
noteworthy: (1) quark interchange contributes about the same amount 
as meson exchanges to the $\bar DN$ $s$-wave phase shifts, and (2) 
among the meson exchanges processes, scalar ($\sigma$) and vector 
($\omega$,$\rho$) are the most important contributors -- single 
$\Lambda_c$, $\Sigma_c$ baryon-exchange diagrams and higher-order 
box diagrams involving $\bar D^* N$, $\bar D\Delta$, and $\bar D^* \Delta$
intermediate states contribute very little. Recall that single-pion
exchange is absent in this reaction. The same model was also used to
examine the possibility to extract information on the $DN$ and $\bar
DN$ interactions in an antiproton annihilation on the
deuteron~\cite{Haidenbauer:2008ff}.

The fact that quark-interchange plays a prominent role in the $\bar
D N$ reaction is very significant to the quest of experimental
signals of chiral symmetry restoration via changes in the
interactions of $\bar D$ mesons in matter. As~said, such changes
would be driven at the microscopic level by modifications of the
properties of the light $u$ and $d$ constituent quarks. However,
within a NRQM there is no direct way to link D$\chi$SB in medium and
the effective hadron-hadron interactions, since constituent quark
masses and microscopic interactions at the quark level are specified
independently in the model~\cite{Close:1979bt}. Also, any temperature 
or density dependence on constituent quark masses has to be 
postulated in an {\em ad hoc} manner~\cite{{Henley:1989vi},{Henley:1989wm}}
to account for effects of D$\chi$SB restoration. As a step 
to remedy this limitation of the NRQM, in the present paper we use a model 
that realizes D$\chi$SB in a way that the same microscopic in\-te\-raction 
that drives D$\chi$SB also confines the quarks and antiquarks into 
color singlet hadronic states, and in addition is the source of the 
hadron-hadron interaction. The model is defined by a microscopic 
Hamiltonian inspired in the QCD Hamiltonian in Coulomb gauge, in that 
an infrared divergent interaction models the full non-Abelian color
Coulomb kernel in QCD and leads to hadron bound states that are
color singlets -- the Hamiltonian also contains an infrared finite
interaction to model transverse-gluon interactions and leads to,
among other effects, hyperfine splittings of hadrons
masses~\cite{{Szczepaniak:2000bi},{Szczepaniak:2001rg},{LlanesEstrada:2004wr}}.
We implement an approximation scheme that allows to calculate with
little computational effort variational hadron wave functions and
effective hadron-hadron interactions that can be iterated in a
Lippmann-Schwinger equation to calculate phase shifts and
cross-sections. An early calculation of the $KN$ interaction within
such a  model, but using a confining interaction only was performed
in Ref.~\cite{Bicudo:1995kq}.

The paper is organizes as follows. In the next Section we present
the microscopic quark-antiquark Hamiltonian of the model. We discuss
D$\chi$SB in the context of two models for the infrared divergent
potentials that mimic full non-Abelian color Coulomb kernel in QCD,
and obtain numerical solutions of the constituent quark mass
function for different current quark masses. In Section~\ref{sec:QM}
we discuss a calculation scheme for deriving effective low-energy
hadron-hadron interactions within the context of the Hamiltonian of
the model. A low-momentum expansion of the quark mass function is
used to obtain va\-ria\-tional meson and baryon wave functions and
explicit, ana\-ly\-ti\-cal expressions for the effective
meson-baryon potential. In Section~\ref{sec:num} we present
numerical results for phase shifts and cross sections for the $K^+N$
and $K^0 N$ and $\bar D^0N$ and $D^- N$ reactions at low energies.
Initially we use the short-ranged quark-interchange potential
derived within the model and add potentials from one-meson exchanges
to fit experimental $s-$wave phase shifts of the $K^+N$ and $K^0 N$
reactions. Next, without introducing new parameters, we present the
predictions of the model for the $\bar D^0N$ and $D^- N$ reactions.
Our conclusions and perspectives are presented in
Section~\ref{sec:concl}. The paper includes one Appendix, that 
presents the meson Lagrangians and respective one-meson exchange
potentials.

\section{Microscopic Hamiltonian and the constituent quark mass function}
\label{sec:H}

The Hamiltonian of the model is given as
\beq
H = H_0 + H_{int},
\label{H}
\eeq
where $H_0$ and $H_{int}$ are given in terms of a quark field operator
$\Psi(\x)$ as
\beq
H_0 = \int d\x \, \Psi^\dagger(\x) (-i\balpha\cdot\bnabla + \beta m)\Psi(\x),
\label{H0}
\eeq
and
\bea
H_{int} &=& - \, \frac 1 2 \int d\x\, d\y \, \rho^a(\x) \, V_C(|\x-\y|) \, \rho^a(\y)
\nn \\
&& + \, \frac 1 2\,\int d\x\,d\y \, J^a_i(\x)\, D^{ij}(|\x-\y|) \, J^a_j(\y) .
\label{Hint}
\eea
In the above $m$ is the current-quark mass matrix of the light~$l=(u,d)$, strange~$s$,
and charm~$c$ quarks:
\beq
m = \left(\begin{array}{cccc}m_u & 0 & 0 & 0 \\ 0 & m_d & 0 & 0 \\ 0 & 0 & m_s & 0
\\0 & 0 & 0 & m_c\end{array}\right),
\label{mass-matrix}
\eeq
and $\rho^a(\x)$ is color charge density
\beq
\rho^a(\x) = \Psi^\dagger(\x)\,T^a\,\Psi(x) ,
\label{rho}
\eeq
$J^a_i(\x)$ is the color current density
\beq
J^a_i(\x) = \Psi^\dagger(\x)\,T^a\alpha_i\,\Psi(\x),
\label{J}
\eeq
with $T^a = \lambda^a/2$, where $\lambda^a$ are the $SU(3)$ Gell-Mann matrices.
$V_C$ and $D^{ij}$ are the effective Coulomb and transverse-gluon interactions;
the transversity of $D^{ij}$ implies
\beq
D^{ij}(|\x - \y|) =
\left( \delta^{ij} - \frac{\nabla^i \nabla^j}{\nabla^2}\right) D_{T}(|\x - \y|) .
\label{Dij}
\eeq
The problem of D$\chi$SB with such an Hamiltonian has been discussed in the 
literature since long time in Bardeen-Cooper-Schriefer (BCS) mean-field type of 
approaches, via Bogoliubov-Valatin transformations or Dyson-Schwinger equations 
in the rainbow approximation~\cite{{Finger:1981gm},{Amer:1983qa},{LeYaouanc:1983it},
{Le Yaouanc:1983iy},{Adler:1984ri},{Kocic:1985uq},{Bicudo:1989sh},{Bicudo:1991kz},
{Mishra:1992we},{Mishra:1994ai},{LlanesEstrada:1999uh},{Bicudo:2001cg},
{LlanesEstrada:2001kr},{Ligterink:2003hd},{Wagenbrunn:2007ie}}.
For our purposes in the present paper, it is more convenient to follow the logic of
the Bogoliubov-Valatin transformations. In this approach, D$\chi$SB is characterized by
a momentum-dependent {\em constituent-quark} mass function $M(k)$, so that the quark
field operator of a given color and flavor can be expanded~as
\begin{equation}
\hspace*{-0.05cm}\Psi(\x) = \sum_s \int \frac{d\k}{(2\pi)^3}
[u_s(\k)q_s(\k) + v_s(\k)\qb^\dag_s(-\k)]e^{i\k\cdot \x},
\label{psi-exp}
\end{equation}
where $u_s(\k)$ and $v_s(\k)$ are Dirac spinors:
\bea
u_s(\k) &=& \sqrt{\frac{E(k) + M(k)}{2E(k)}}
\left(\begin{array}{c}
1
\\[0.3true cm]
 \,\frac{\bsigma\cdot \k}{E(k) + M(k)} \end{array}
\right) \chi_s,
\label{u} \\[0.5true cm]
v_s(\k) &=& \sqrt{\frac{E(k) + M(k)}{2E(k)}}
\left(\begin{array}{c}
-  \frac{\bsigma\cdot\k}{E(k) + M(k)}
\\
[0.3true cm] 1 \end{array}\right)
\chi^c_s,
\label{v}
\eea
with $E(k) = [k^2 + M^2(k)]^{1/2}$, $\chi_s$ is a Pauli spinor, $\chi^c_s =
-i\,\sigma^2\chi^\ast_s$, and $q^\dag_s(\k)$, $\qb^\dag_s(\k)$, $q_s(\k)$,
and $\qb_s(\k)$ are creation and annihilation operators of {\em constituent
quarks}; $q_s(\k)$ and $\qb_s(\k)$ annihilate the vacuum state $|\Omega\rangle$:
\begin{equation}
\label{cond}
q_{s}(\bm{k})|\Omega\rangle=0,
\hspace{1.0cm}
\bar{q}_{s}(\bm{k})|\Omega\rangle=0 .
\end{equation}
For $m = 0$, the Hamiltonian is chirally symmetric, but $|\Omega\rangle$
is not symmetric, $\langle \Omega |\bar\Psi \Psi |\Omega \rangle \neq 0$.

Substituting in Eqs.~(\ref{H0}) and (\ref{Hint}) the expansion of $\Psi$,
Eq.~(\ref{psi-exp}), and rewriting $H$ in Wick-contracted form, one
obtains an expression for $H$ that can be written as a sum of three
parts:
\beq
H = {\cal E} + H_2 + H_4,
\label{HWick}
\eeq
where ${\cal E}$ is the c-number vacuum energy, and $H_2$ and $H_4$ are
normal-ordered operators, respectively quadratic and quartic in the
creation and annihilation operators. The mass function $M(k)$ is determined
by demanding that $H_2$ be diagonal in the quark operators. This leads
to the {\em gap equation} for the constituent quark mass function $M_f(k)$
of flavor $f$:
\begin{eqnarray}
\label{gap}
M_f(k) &=& m_f + \frac{2}{3}\int\frac{d\q}{(2\pi)^{3}}\,
\Bigl[F^{(1)}_f(\bm{k},\bm{q})\, V_{C}(|\bm{k}-\bm{q}|)
\nonumber \\
&& + \,  2\,G^{(1)}_f(\bm{k},\bm{q})\,D_{T}(|\bm{k}-\bm{q}|)\Bigr],
\end{eqnarray}
where $m_f$ is the current quark mass and
\bea
F^{(1)}_f(\k,\q) &=& \frac{M_f(q)}{E_f(q)} -
\frac{M_f(k)}{E_f(q)} \frac{q}{k}\hat{\k}\cdot\hat{\q},
\label{f1} \\[0.3true cm]
G^{(1)}_f(\k,\q) &=& \frac{M_f(q)}{E_f(q)} \nn \\
&& +\, \frac{M_f(k)}{E_f(q)}\frac{q}{k}\frac{(\bm{k}\cdot\bm{q}-k^{2})
(\bm{k}\cdot{\bm{q}} - q^{2})}{k q|\bm{k}-\bm{q}|^{2}} ,
\label{g1}
\eea
and $V_C(k)$ and $D_T(k)$ are the Fourier transforms of $V_C(|\x|)$
and $D_T(|\x|)$. The quadratic Hamiltonian $H_2$ is given by
\beq
H_2 = \sum_{s,f} \int d\k  \,\varepsilon_f(k)
\left[q^{\dagger}_{sf}(\bm{k})q_{sf}(\bm{k}) +
\bar{q}^{\dagger}_{sf}(\bm{k})\bar{q}_{sf}(\bm{k})\right] ,
\label{H2}
\eeq
where $\varepsilon_f(k)$ is the constituent-quark single-particle
ener\-gy of flavor $f$, given by
\begin{eqnarray}
\varepsilon_f(k) &=&\frac{k^{2}+m_f M_f(k)}{E_f(k)} \nn \\
&& + \frac{2}{3}\int\frac{d\q}{(2\pi)^{3}} \Bigl[F^{(2)}_f(\bm{k},\bm{q}) \,
V_{C}(|\bm{k}-\bm{q}|)
\nonumber \\
&& + \, 2 G^{(2)}_f(\bm{k},\bm{q}) \, D_{T}(|\bm{k}-\bm{q}|) \Bigr] ,
\label{epsilon}
\end{eqnarray}
with
\beq
F^{(2)}_f(\bm{k},\bm{q}) = \frac{M_f(k)}{E_f(k)}\frac{M_f(q)}{E_f(q)} +
\frac{\k\cdot\q}{E_f(k) E_f(q)} ,
\label{F2}
\eeq
and
\beq
G^{(2)}_f(\bm{k},\bm{q}) = \frac{M_f(k)}{E_f(k)}\frac{M_f(q)}{E_f(q)}
+\frac{(\bm{k}\cdot\bm{q}-k^{2})(\bm{k}\cdot\bm{q}-q^{2})}{E_f(k)E_f(q)
|\bm{k}-\bm{q}|^{2}}.
\label{G2}
\eeq
The four-fermion term $H_4$ is simply the normal-ordered form of $H_{int}$:
\beq
H_4 = \, :H_{int}: .
\label{H_4}
\eeq
The color confining feature of the Hamiltonian will be discussed in the next
Section.

To solve the gap equation on needs to specify the interactions $V_C$
and $D_T$. For the confining Coulomb term $V_C$, we use two
analytical forms, that we name Model~1 and Model~2, to assess the
sensitivity of results with res\-pect to $V_C$. Model~1 is a
parametrization of the lattice simulation of QCD in Coulomb gauge of
Ref.~\cite{Voigt:2008rr}:
\begin{equation}
\label{model1}
V_{C}(k) = \frac{8\pi\sigma_\text{Coul}}{k^4} + \frac{4\pi C}{k^2},
\end{equation}
with $\sigma_\text{Coul} = (552\,{\rm MeV})^2$ and $C=6$. Model~2 for $V_C$
was used in recent studies of glueballs~\cite{Guo:2007sm} and heavy hybrid
quarkonia~\cite{Guo:2008yz}; it is written as
\begin{equation}
\label{model2}
V_{C}(k) = V_{l}(k) + V_{s}(k),
\end{equation}
where
\begin{equation}
V_{l}(k)=\frac{8 \pi \sigma}{k^{4}},
\hspace{.5cm}
V_{s}(k)=\frac{4 \pi \alpha(k)}{k^{2}},
\label{Vl+Vs}
\end{equation}
with
\begin{equation}
\alpha(k) = \frac{4 \pi Z}{\beta^{3/2}\ln^{3/2}
\left(c + k^{2}/\Lambda^2_\text{QCD}\right)} .
\label{alpha}
\end{equation}
The parameters here are $\Lambda_\text{QCD}=250$\,MeV, $Z =5.94$,
$c = 40.68$, and $\beta = 121/12$. For the transverse-gluon
interaction $D_T$ we use
\begin{equation}
D_{T}(k)=-\frac{4\pi\alpha_{T}}{({k}^{2}+m^{2})
\ln^{1.42}(\tau + {k}^{2}/m^{2}_{g})}.
\label{DT}
\end{equation}
This choice is guided by previous studies of spin-hyperfine
splittings of meson masses~\cite{LlanesEstrada:2004wr} using
an Hamiltonian as in the present work. Moreover, the Yukawa
term multiplying the log term is used to conform lattice
results that the gluon propagator in Coulomb gauge is finite
in the infrared~\cite{{Voigt:2008rr},{Nakagawa:2009zf}}. Further
ahead we will discuss the impact of this choice of infrared
behavior on our numerical results. Parameters here are $m_{g}=550$~MeV,
$m =m_{g}/2$, $\tau = 1.05$, and $\alpha_{T} = 0.5$. We use the same
$D_T$ for both models. We note that we could have used a log running
similar to the one in $V_C$, Eq.~(\ref{alpha}), but results would not
change in any significant way.

We have solved the gap equation in Eq.~(\ref{gap}) by ite\-ra\-tion.
The angular integrals can be performed analytically, but special
care must be taken with the strongly-peaked confining term $1/|\k
-\q|^4$ at $\q \approx \k$ in the numerical integration over $q$. There
is no actual divergence here: the terms $M(q)/E(q)$ and $M(k)/E(q)
\, (q/k) \,\hat\k\cdot\hat\q$ in Eq.~(\ref{f1}) cancel exactly when
$\q = \k$, but this cancellation can be problematic in the numerical
integration. This problem can be handled in different
manners, like using a momentum mesh containing node and half-node
points with e.g. $k$ at nodes and $q$ at
half-nodes~\cite{Adler:1984ri}, or introducing a mass parameter
$\mu_{IR}$ such $k^4 \rightarrow (k^2 + \mu^2)^2$ in
Eq.~(\ref{model1}) and varying $\mu_{IR}$ until results become
independent of $\mu_{IR}$~\cite{Alkofer:2005ug}. In the present
paper we use a different method~\cite{adam}; we add a convenient
zero to the gap equation:
\bea
- \int \frac{d\q}{(2\pi)^3} \, \frac{1}{|\k - \q|^4}\, \frac{M(k)}{E(k)} \,
\frac{1}{k}\hat k \cdot \left(\k - \q\right),
\label{zero}
\eea
and rewrite $F^{(1)}(\k,\q)$ in Eq.~(\ref{gap}) as
\bea
F^{(1)}(\k,\q) &\rightarrow & F^{(1)}(\k,\q) - \frac{M(k)}{E(k)} \,
\frac{1}{k}\hat k \cdot \left(\k - \q\right) \nn \\
&=& \left(\frac{M(q)}{E(q)}
- \frac{M(k)}{E(q)} \right) \nn \\
&& \hspace*{-0.2cm} - \, \left(\frac{q}{k} \,\hat\k\cdot\hat\q - 1 \right)
\left(\frac{M(k)}{E(q)}
- \frac{M(k)}{E(k)} \right),
\eea
so that angle-independent and angle-dependent terms vanish independently
when $\q = \k$. This feature makes the numerical cancellation of the
divergences very stable.

\begin{figure}[t]
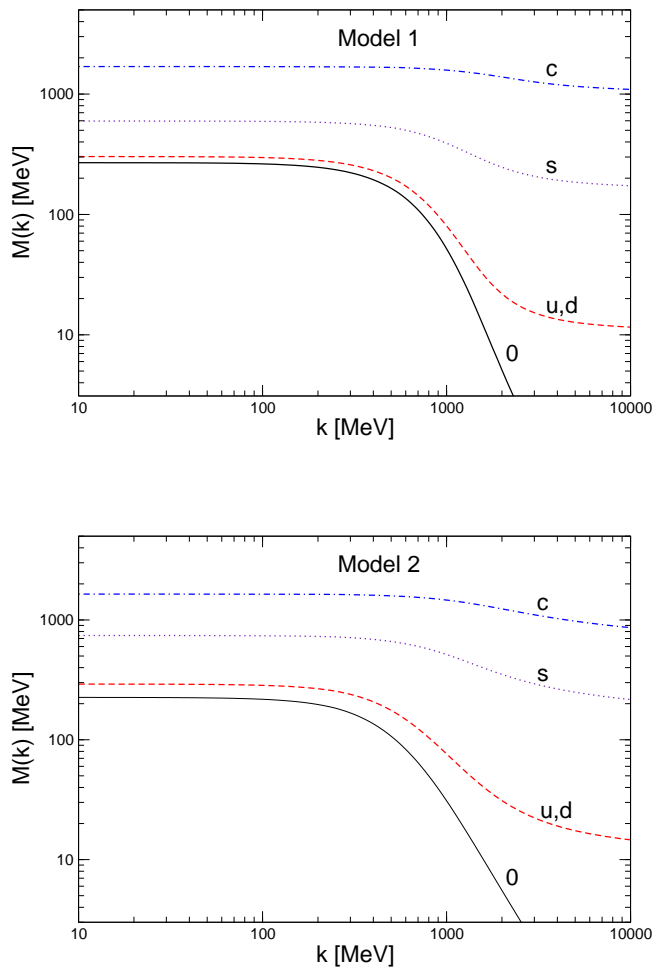

\begin{center}
\includegraphics[scale=0.34]{fig1a_mk1.eps}\\[1.2truecm]
\includegraphics[scale=0.34]{fig1b_mk2.eps}
\end{center}
\vspace{-.3cm}
\caption{Constituent quark mass $M(k)$ as a function of the
momentum for different values of current-quark masses.}
\label{fig:gap}
\end{figure}

In Fig.~\ref{fig:gap} we present solutions $M(k)$ of the gap equation for
different values of current-quark masses. The dramatic effect of D$\chi$SB
mass-generation is seen clearly in the figure: for the fictitious limit
of a zero current-quark mass (solid line), the mass function $M(k)$ acquires
a si\-za\-ble value in the infrared, $M(k=0) \approx 250$~MeV. In the ultraviolet,
the mass function runs logarithmically with $k$, in a manner dictated by the
running of the microscopic interactions. Since the model interactions used here
have a different ultraviolet running from the one dictated by perturbative
QCD, the logarithm running of the quark-mass function must be different; the
interactions here fall off faster than those in QCD. One consequence of this
is that the momentum integral over the trace of the quark propagator, which
gives the quark condensate, does not run and is ultraviolet finite in the 
present case; we obtain
$(-\langle \bar\Psi\Psi\rangle)^{1/3}~=~280$~MeV. For nonzero current-quark
masses, Fig.~\ref{fig:gap} shows that the effect of mass ge\-ne\-ra\-tion diminishes
as the value of $m_f$ increases -- here $m_u = m_d = 10$~MeV, $m_s = 150$~MeV
in both models, and $m_c = 950$~MeV in Model~1 and $m_c = 600$~MeV in Model~2.
On the logarithm scale, one sees that the mass function varies substantially
in the ultraviolet, from $k \simeq 3 \Lambda_{\rm QCD} = 750$~MeV onwards.
It is worth noticing that although the momentum dependence of the mass function
charm quark is much less dramatic than for light quarks, there is still a
significant dressing effect, $M_c(0)/m_c \sim 1.5 - 3$, similarly to what is
found with covariant Dyson-Schwinger equations~\cite{{Nguyen:2010yh},
{Roberts:2007jh},{Alkofer:2003jj}}.

Finally, we comment on the impact of the momentum dependence of $D_T(k)$ in
the infrared on the mass function. In the complete absence of $D_T$, using
only $V_C$ with the parameters above, $M(0)$ would be of the order of
$100$~MeV in the chiral limit. On the other hand, using a $D_T(k)$ that
vanishes at $k=0$, like in the form of a Gribov formula~\cite{gribov},
would give $M(0) \sim 200$~MeV in the chiral limit. This is in line with the
recent finding of Ref.~\cite{Pak:2011wu}, which studies D$\chi$SB in a framework
based on a quark wave functional determined by the variational principle using
an ansatz which goes beyond the BCS-type of wave functionals. 

\section{Baryon-meson interaction}
\label{sec:QM}

In this Section we set up a calculation scheme for de\-ri\-ving
effective low-energy hadron-hadron interactions with the Hamiltonian
discussed above. We seek a scheme that makes contact with
traditional quark-model calculations and can be systematically
improved with some computational effort. Since the early 80's, the
great majority of calculations of hadron-hadron scattering
observables in the quark-model are based on methods to handle
cluster dynamics adapted from nuclear or atomic physics -- for
reviews on methodology and fairly complete lists of re\-fe\-ren\-ces
see Refs.~\cite{{Myhrer:1987af},{Hadjimichef:1998rx},{Hadjimichef:2000en}}.
Such methods require a mi\-cros\-co\-pic Hamiltonian and hadron bound
state wave functions given in terms of quark degrees of freedom.
Within the context of the model discussed in the previous Section,
the starting point is the microscopic Hamiltonian of Eq.~(\ref{H}),
with quark field operators given in terms of the constituent quark
mass function $M(k)$ obtained from the numerical solution of the gap
equation in Eq.~(\ref{gap}).

The sector of the Hamiltonian relevant for the elastic meson-baryon interaction can
be written in a compact notation as
\begin{eqnarray}
H_{q\bar q} &=& \varepsilon(\mu) \, q^{\dag}_{\mu}q_{\mu}
+ \varepsilon(\nu)\,\bar{q}^{\dag}_{\nu} \bar{q}_{\nu}
+ \frac{1}{2}V_{qq}(\mu\nu;\sigma\rho) \, q^{\dag}_{\mu}
q^{\dag}_{\nu}q_{\rho}q_{\sigma}
\nonumber\\
&& \hspace*{-0.7cm} + \frac{1}{2}V_{\bar{q}\bar{q}}(\mu\nu;\sigma\rho)\,
\bar{q}^{\dag}_{\mu}\bar{q}^{\dag}_{\nu}\bar{q}_{\rho}\bar{q}_{\sigma} +
V_{q\bar{q}}(\mu\nu;\sigma\rho)\,q^{\dag}_{\mu}
\bar{q}^{\dag}_{\nu}\bar{q}_{\rho}q_{\sigma},
\label{Hqq}
\end{eqnarray}
where the indices $\mu$, $\nu$, $\rho$, and $\sigma$ denote collectively the
quantum numbers (orbital, color, spin, flavor) of quarks and antiquarks.
The first two terms in Eq.~(\ref{Hqq}) are the quark and antiquark
single-particle self-energies coming from Eq.~(\ref{H2}),
and $V_{qq}$, $V_{q\bar{q}}$, and $V_{\bar q\bar{q}}$ are respectively the
quark-quark, quark-antiquark and antiquark-antiquark interactions from $H_4$
in Eq.~(\ref{H_4}). The one-baryon state in the BCS approximation can be
written in the schematic notation as~\cite{{Bicudo:1991kz},{LlanesEstrada:2001kr},
{Bicudo:2001cg}}
\begin{equation}
|a\rangle = B^{\dag}_{a}|\Omega\rangle = \frac{1}{\sqrt{3!}} \,
\psi^{\mu_{1}\mu_{2}\mu_{3}}_{a} \, q^{\dag}_{\mu_{1}}
q^{\dag}_{\mu_{2}}q^{\dag}_{\mu_{3}}|\Omega\rangle,
\label{baryon}
\end{equation}
with $\psi^{\mu_{1}\mu_{2}\mu_{3}}_{a}$ being the Fock-space
amplitude, with $a$ denoting the (orbital, spin, flavor) quantum numbers
of the baryon. Likewise, the one-meson states is written as
\begin{equation}
|a\rangle = M^{\dag}_{a}|\Omega\rangle =
\phi_{a }^{\mu\nu}q^{\dag}_{\mu}\bar{q}^{\dag}_{\nu}|\Omega\rangle ,
\label{meson}
\end{equation}
where $\phi_{a }^{\mu\nu}$ is the corresponding Fock-space amplitude,
with $a$ representing the quantum numbers of the meson. The Fock-space
amplitudes $\psi$ and $\phi$ can be obtained by solving a
Salpeter-type of equation~\cite{{Bicudo:1991kz},{LlanesEstrada:2001kr},
{Bicudo:2001cg}}.

Given a mi\-cros\-co\-pic Hamiltonian $H_{q\bar q}$ and hadronic
states $|a\rangle$ as in Eqs.~(\ref{Hqq})-(\ref{meson}), an effective
low-energy meson-baryon $V_{MB}(ab,cd)$ potential for
the process $\text{Meson}(a) + \text{Baryon}(b) \rightarrow \text{Meson}(c) +
\text{Baryon}(d)$ can be written as~\cite{Hadjimichef:1998rx}
\begin{eqnarray}
V_{MB}(ab,cd) &=&-\,3\,\phi_{c}^{*\mu\nu_{1}}
\psi_{d}^{*\nu\mu_{2}\mu_{3}}
V_{qq}(\mu\nu;\sigma\rho)\phi_{a}^{\rho\nu_{1}}
\psi_{b}^{\sigma\mu_{2}\mu_{3}}\nonumber\\[0.17cm]
&&\hspace*{-0.7cm} -\,3\,\phi_{c}^{*\sigma\rho}\psi_{d}^{*\mu_{1}\mu_{2}\mu_{3}}
V_{q\bar{q}}(\mu\nu;\sigma\rho)\phi_{a}^{\mu_{1}\rho}
\psi_{b}^{\mu\mu_{2}\mu_{3}}\nonumber\\[0.17cm]
&&\hspace*{-0.7cm} -\,6\,\phi_{c}^{*\mu_{1}\nu_{1}}\psi_{d}^{*\nu\mu\mu_{3}}
V_{qq}(\mu\nu;\sigma\rho)\phi_{a}^{\rho\nu_{1}}
\psi_{b}^{\mu_{1}\sigma\mu_{3}}\nonumber\\[0.17cm]
&&\hspace*{-0.7cm} -\,6\,\phi_{c}^{*\nu_{1}\nu}\psi_{d}^{*\nu_{1}\mu\mu_{3}}
V_{q\bar{q}}(\mu\nu;\sigma\rho)\phi_{a}^{\nu_{1}\rho}
\psi_{b}^{\mu_{1}\sigma\mu_{3}} ,
\label{V_MB}
\end{eqnarray}
where $\phi^{\mu \nu}_a$, $\psi^{\mu\nu\sigma}_b$, $\cdots$ are the
meson and baryon Fock-space amplitudes of Eqs.~(\ref{baryon}) and
(\ref{meson}), and $V_{qq}$, $V_{q\bar{q}}$, and $V_{\bar q\bar{q}}$
are respectively the quark-quark, quark-antiquark and antiquark-antiquark
interactions in $H_{q\bar q}$ of Eq.~(\ref{Hqq}). The expression in
Eq.~(\ref{V_MB}) is completely general, in that it is valid for any
low-energy meson-baryon process for which the baryon and meson Fock-space
states and $H_{q\bar q}$ are as given above. It can be iterated in a Lippmann-Schwinger
equation to obtain scattering phase shifts and cross sections -- for details, see
Ref.~\cite{Hadjimichef:1998rx}. There is, however, one difficulty:
$V_{MB}$ involves multidimensional integrals over internal
quark and antiquark momenta of products of baryon and meson wave functions
and products of Dirac spinors $u(\k)$ and $v(\k)$ that depend on the quark
mass function $M(k)$. Although this is not a major difficulty, an approximation
can be made noting that the bound-state amplitudes $\psi$ and $\phi$
are expected to fall off very fast in momentum space for momenta larger than
the inverse size of the hadron, and therefore only low-momentum quark and antiquark
processes contribute in the multidimensional integrals.
In~view of this, and the fact that $M(k)$ changes considerably only at
large momenta, a na\-tu\-ral approximation scheme to simplify the calculations
without sacrificing the low-energy content of the effective interaction,
is to retain the first few terms in the low momentum expansion for the
mass function $M(k)$:
\begin{equation}
M(k) = M + M'(0)\, k + \frac{1}{2} M''(0) \,k^2 + \cdots ,
\label{Mk-app}
\end{equation}
where $M'(k)= dM(k)/dk$ and $M''(k)= d^2M(k)/dk^2$. In particular, retaining
terms up to ${\cal O}(k^2/M^2)$ in the expansion, it is not difficult to
show that the Dirac spinors $u_s(\k)$ and $v_s(\k)$ in Eqs.~(\ref{u}) and
(\ref{v}) become
\begin{eqnarray}
u_{s}(\k) &=&
\left(\begin{array}{c} 1 - \frac{k^2}{8M^2} \\[0.1true cm]
\frac{{\bm\sigma}\cdot{\k}}{2M}
\end{array}\right)\,\chi_{s},  
\label{u-app}
\\[0.5true cm]
v_{s}(\k) &=& \left(\begin{array}{c}
- \frac{\bm{\sigma}\cdot\k}{2M} \\
[0.1true cm] 1 - \frac{k^2}{8M^2}
\end{array}\right)\,\chi^{c}_{s}.
\label{v-app}
\end{eqnarray}
Using these spinors in Eq.~(\ref{Hint}), the expressions one obtains
for $V_{qq}$, $V_{q\bar{q}}$, and $V_{\bar q\bar{q}}$ are very similar
to those of the Fermi-Breit expansion of the OGE
interaction~\cite{De Rujula:1975ge}. There is, however, one important
difference here: while for the OGE one has $V_C(k) = D_T(k) \approx 1/k^2$,
in the present case $V_C(k)$ and $D_T(k)$ are different and represent
very different physics; $V_C$ is a confining interaction and $D_T$ is a (static)
transverse gluon interaction.

The evaluation of multidimensional integrals can be further simplified using
the variational method of Refs.~\cite{{Bicudo:1991kz},{Bicudo:2001cg}} with
Gaussian ans\"atze for the bound-state amplitudes~$\psi$~and~$\phi$, instead
of solving numerically Salpeter-type of equations. Specifically:

\begin{eqnarray} 
\hspace*{-0.4cm}\psi_{\P}(\bm{k}_{1},\bm{k}_{2},\bm{k}_{3}) &=&
\delta({\bm P} - {\bm k}_1 - {\bm k}_2 - {\bm k}_3) \nonumber \\[0.17cm]
 &\times& \,  \left(\frac{3}{\pi^{2}\alpha^{4}}\right)^{3/4}
e^{-\sum^3_{i=1} (\,\bm{k}_{i}-\bm{P}/3\,)/2\alpha^2 },
\label{psi}
\end{eqnarray}

\begin{eqnarray}
\phi_{\P}(\bm{k}_{1},\bm{k}_{2}) &=& \delta({\bm P}
- {\bm k}_1 - {\bm k}_2) \nonumber \\[0.17cm]
 &\times& \, \left(\frac{1}{\pi\beta^{2}}\right)^{3/4}
e^{ - (M_{1}\bm{k}_{1} - M_{2}\bm{k}_2)^2/8\beta^2 },
\label{phi}
\end{eqnarray}


\noindent
where $\alpha$ and~$\beta$ are variational parameters, ${\bm P}$ is the
center-of-mass momentum of the hadrons and 
\begin{eqnarray}
M_{1} = \frac{2M_{\bar{q}}}{(M_{q}+M_{\bar{q}})}, 
\hspace{0.5cm}
M_{2} = \frac{2M_{q}}{(M_{q}+M_{\bar{q}})},
\end{eqnarray}
with $M_q$ and $M_{\bar{q}}$ being the
zero-momentum constituent-quark masses. The va\-ria\-tional parameters are
determined by mi\-ni\-mi\-zing the hadron masses:
\begin{equation}
M_{a}=\frac{\langle a|\,(H_{2}+H_{4})\,|a'\rangle}
{\langle a| a'\rangle}{\Bigg|}_{\bm{P}_{a}=\bm{P}_{a'}=0},
\label{Ma}
\end{equation}
where we left out the constant $\cal E$, defined in Eq.~(\ref{HWick}), which
cancels in the hadron mass differences -- see Table~I.

In this work we consider the elastic scattering of the
pseudoscalar mesons $(K^+,K^0)$ and $(\bar D^0,D^-)$ off nu\-cleons,
with both the mesons and nucleons in their ground states. Using the
Gaussian forms for the amplitudes $\psi$ and $\phi$, Eqs.~(\ref{psi}) and 
(\ref{phi}), one obtains for the nucleon mass, $m_N$, and for the pseudoscalar
meson mass, $m_P$, the following expressions:

\begin{widetext}
\begin{eqnarray}
M_N &=& 3\left(\frac{3}{\pi\alpha^2}\right)^{3/2}
\int d\k\left[\frac{k^2}{E_l(k)} +
m_l \, \frac{M_l(k)}{E_l(k)}\right] e^{-3 k^2/\alpha^2} \nn \\
&& + \, \left(\frac{3}{2\pi\alpha^2}\right)^{3/2}
\int\frac{d\k d\q}{(2\pi)^{3}}
\Biggr[2 F^{(2)}_l(\k,\q)
+ 3 \, C_N \, e^{-(\k-\q)^2/2\alpha^2}\Biggr]
V_{C}(|\k-\q|) \, e^{-3k^2/2\alpha^2} \nn \\
&& + \,\left(\frac{3}{2\pi\alpha^2}\right)^{3/2}
\int\frac{d\k d\q}{(2\pi)^3}
\Biggl[4 G^{(2)}_l(\k,\q) - \frac{(\k-\q)^2}{3 M^2_l}
e^{-(\k - \q)^2/2\alpha^2}{\Bigg]}
D_{T}(|\k-\q|) \, e^{- k^2/\alpha^2},
\label{mN}
\\[0.5true cm]
M_P &=& \left(\frac{1}{\pi\beta^{2}}\right)^{3/2} \int d\k
\left[\frac{k^2}{E_l(k)} + m_l \,\frac{M_l(k)}{E_l(k)} +
\frac{k^2}{E_h(k)} + m_h \, \frac{M_h(k)}{E_h(k)}\right]
e^{-k^2/\beta^2}
\nonumber \\
&& + \, \left(\frac{1}{\pi\beta^2}\right)^{3/2}
\int \frac{d\k d\q}{(2\pi)^3}
\Biggl\{\frac{2}{3}\left[F^{(2)}_l(\k,\q) + F^{(2)}_h(\k,\q)\right]
+ \, C_P \,
 e^{-(\k-\q)^2/2\beta^2} \Biggr\}
V_{C}(| \k - \q |) \, e^{-k^2/\beta^2}
\nonumber \\
&& + \, \frac{4}{3}\left(\frac{1}{\pi\beta^2}\right)^{3/2}
\int \frac{d\k d\q }{(2\pi)^3}
\Biggl[G^{(2)}_l(\k,\q) + G^{(2)}_h(\k,\q)
+ \frac{1}{2}\frac{(\k-\q)^2}{M_l M_h}
e^{-(\k - \q)^2/2\beta^2} \Biggr]
D_{T}(|\k-\q|) \, e^{-k^2/\beta^2},
\label{mP}
\end{eqnarray}
\end{widetext}

\noindent
where $F^{(2)}_{l,h}(\k,\q)$ and $G^{(2)}_{l,h}(\k,\q)$ are given in 
Eqs.~(\ref{F2}) and
(\ref{G2}) and the indices $l$ and $h$ refer to light and heavy flavors,
$l=(u,d)$ and $h=(s,c)$. The matrix elements of the color matrices, $C_N =
\langle N| T^a T^a |N \rangle$ and $C_P = \langle P| T^a (-T^{a})^T |P \rangle$,
in the terms proportional to the Coulomb potential $V_C$ -- that come from $H_4$
-- are written in a way to emphasize the color-confinement feature of the
model~\cite{Bicudo:1991kz}: for $\k = \q$, these terms are divergent,
{\em unless} the color matrix elements are such that the cor\-responding
expressions vanish. The terms $F^{(2)}_{l,h}(\k,\q)$ come from $H_2$
-- which is diagonal in color. For baryon and meson {\em color singlet states},
the contributions from $H_2$ and $H_4$ cancel exactly for $\k = \q$, since:
\bea
C_N &=& \frac{\varepsilon^{c_1 c_2 c_3}\varepsilon^{c'_1 c'_2 c_3}}{3!}
\, (T^a)^{c_1 c'_1} (T^a)^{c_2 c'_2} = - \frac{2}{3},
\label{Ncolor}
\eea

\bea
C_P &=& \frac{\delta^{c_1 c_2}\, \delta^{c'_1 c'_2}  }{3}\,
(T^a)^{c_1 c'_1} (-T^a)^{c'_2 c_2}
= - \frac{4}{3},
\label{Pcolor}
\eea
and
\bea
\lim_{\k \rightarrow \q} F^{(2)}_{l,h}(\k,\q) = 1.
\label{lim}
\eea
Such a cancellation of infrared divergences also plays an important role
in context of a conjectured~\cite{McLerran:2007qj} new high-density phase
of matter composed of confined but chirally symmetric
hadrons~\cite{{Glozman:2008fk},{Glozman:2009sa}}.

\begin{figure}[htbp]
\begin{center}
\resizebox{8.5cm}{!}{\includegraphics{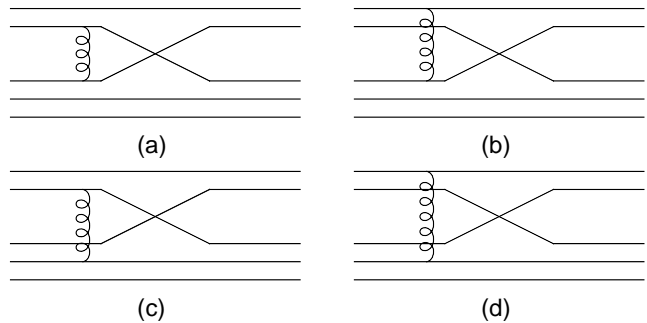}}
\caption{Pictorial representation of the quark-interchange
processes that contribute to a meson-baryon interaction. The curly
lines represent the interactions $V_C$ and $D_T$. }
\label{fig:qg-inter}
\end{center}
\end{figure}

Next, we obtain an explicit expression for the effective meson-nucleon interaction $V_{MB}$, 
given generically in Eq.~(\ref{V_MB}). This effective interaction is generated by a
quark-interchange mechanism. The use of Gaussian wave functions is very helpful
for getting a $V_{MB}$ in closed form~\cite{Hadjimichef:1998rx}: it can be written as a 
sum of four contributions (see Eqs.~(9) and (10) of Ref.~\cite{Haidenbauer:2007jq}), 
each contribution corresponding to a quark-interchange diagram shown in
Fig.~\ref{fig:qg-inter}:
\beq
V_{MB}(\p,\p') = \frac{1}{2} \sum_{i=1}^{4}\omega_{i}
\left[\,V_{i}(\p,\p') + V_{i}(\p',\p)\,\right],
\label{V_MB-sum}
\eeq
where $\p$ and $\p'$ are the initial and final center-of-mass
momenta, and the $V_i(\p,\p')$ are given by
\bea
V_{i}(\bm{p},\bm{p}')&=&\left[\,\frac{3g}{(3+2g)\pi\alpha^{2}}\,\right]
e^{-\,a_{i}p^{2} - b_{i}p'^{2} + c_{i}\bm{p}\cdot\bm{p}'\,}\nonumber\\
&\times& \int\frac{d\q}{(2\pi)^{3}}\,v(q)\,e^{-d_{i}q^{2}
+ \bm{e}_{i}\cdot\bm{q}} ,
\label{Vi}
\eea
where the $a_{i}$, $b_{i}$, $c_{i}$, $d_{i}$, and $e_{i}$ involve the
hadron sizes $\alpha$ and $\beta$ and the constituent quark masses $M_{u}$,
$M_{s}$ and $M_{c}$, and $g = (\alpha/\beta)^2$. Since we are using the same
Fock-space amplitudes $\psi$~and~$\phi$ as those used in Ref.~\cite{Haidenbauer:2007jq},
the expressions for $a_{i}$, $b_{i}$, $\cdots$ are the same as there. It~is, ho\-we\-ver,
important to note that the essential difference here as compared to
Ref.~\cite{Haidenbauer:2007jq} is that the constituent quark masses $M_u$, $M_s$,
and $M_c$, the width parameters $\alpha$ and $\beta$, and the meson-baryon interaction
are all derived from the same microscopic quark-gluon Hamiltonian. Another very important
difference here is $v(q)$: while in Ref.~\cite{Haidenbauer:2007jq} it comes from the OGE,
here $v(q) = V_C(q)$ for the spin-independent interaction and $v(q) = 2q^2/(3M_l M')
\, D_T(q)$, for the spin-spin interaction, with $M'= M_l$ for diagrams~$(a)$ and $(c)$,
and $M' = M_h$ for diagrams $(b)$ and $(d)$. The $\omega_i$ are coefficients that come
from the sum over quark color-spin-flavor indices and combinatorial factors, whose
va\-lues for the $\bar{D}N$ interaction are given in Table~3 of
Ref.~\cite{Haidenbauer:2007jq}. The corresponding coefficients for the $KN$ reaction
can be extracted from that table by identifying $\bar{D}^{0}$ with
$K^{+}$ and $D^{-}$ with $K^{0}$. Note that $V_{MB}(\p,\p')$ is symmetric under
interchange of $\p$ and $\p'$ and therefore possess post-prior
symmetry~\cite{{Wong:1999zb},{Hadjimichef:1998rx}}, a feature that is not always satisfied
when using composite wave functions that {\em are not} exact eigenstates of
the microscopic Hamiltonian~\cite{Schiff}.
As mentioned earlier, the quark-interchange mechanism leads to an
effective meson-baryon potential that is of very short range. It depends
on the overlap of the hadron wave-functions and contributes
mostly to $s$-waves. As shown in Ref.~\cite{Hadjimichef:2002xe},
in order to describe experimental data for the $K^+N$ and $K^0N$
reactions, light-meson exchange processes are required to account
for medium- and long-ranged components of the force. Quark-interchange
generated by OGE accounts only to roughly 50~\% to the experimental
$s$-wave phase shifts and vector ($\omega$ and $\rho$) and scalar exchanges
are crucial for the correct description of this wave as well as higher partial
waves~\cite{Hadjimichef:2002xe}. Moreover, in order to describe the correct
isospin dependence of the data, it is essential to include the exchange of
the scalar-isovector $a_0$ meson~\cite{Hadjimichef:2002xe}. With these facts
in mind, we follow Ref.~\cite{Haidenbauer:2007jq} and include meson-exchanges
in the $\bar D N$ system. As in that re\-fe\-rence, we parameterize correlated
$\pi\pi$ exchange in terms of a single $\sigma$-meson exchange -- this is
not a bad appro\-ximation for the $I=0$ channel, but for $I=1$ channel
it underestimates the total strength by 50\%~\cite{Haidenbauer:2007jq}.
In~Appendix~\ref{app:mex} we present the effective Lagrangians densities
and corresponding one-meson-exchange amplitudes.

%
\section{Numerical results for phase shifts and cross sections}
\label{sec:num}

As mentioned previously, not much is known experimentally about
the $\bar D N$ interaction at low energies. In view of this and in
order to have a comparison standard, we also consider within the
same model, wi\-thout chan\-ging any parameters besides the current
quark masses, the $K^+N$ and $K^0 N$ elastic processes, for which
there are experimental data. It is important to rei\-terate that
the effective short-range meson-baryon potential derived from
quark-interchange driven by the mi\-cros\-copic in\-te\-ractions
of Model~1 or Model~2 are very cons\-trained, in the sense that
they are determined by the mi\-cros\-copic in\-te\-rac\-tions $V_C$
and $D_T$ via a chain of intermediate results for hadron
properties driven by D$\chi$SB. That is, the microscopic
interactions $V_C$ and $D_T$ determine the constituent quark
masses $M_l=(M_u,M_d)$ and $M_h = (M_s, M_c)$, and these quark
masses together with $V_C$ and $D_T$ determine the hadron wave
functions. The effective meson-baryon interaction is determined by
all these ingredients si\-mul\-ta\-neously, since it depends
ex\-pli\-citly on the hadron wave functions, the quark masses and
$V_C$ and $D_T$. Such an interdependent chain of results
determining the effective meson-baryon interaction is absent in
the NRQM with OGE, where quarks masses are independent of the
microscopic quark-antiquark Hamiltonian.

Having determined the constituent quarks masses $M_l=(M_u,M_d)$ and 
$M_h = (M_s, M_c)$, the next step is the determination of the variational 
parameters $\alpha$ and $\beta$ of the $\psi$ and $\phi$ wave functions. 
Minimization of $M_N$ and $M_P$ with respect to $\alpha$ and $\beta$ 
leads to the results shown in Table~\ref{tab:sizes}; also shown are the 
parameters used with the OGE model in Refs.~\cite{{Hadjimichef:2002xe},
{Haidenbauer:2007jq}}.

As~expected, because the charm quark is heavier than the strange
quark, the charmed $D$ mesons are smaller objects than the kaons -
recall that the meson root mean square radii are inversely
proportional to $\beta$. One also sees that hadron sizes of Models~1
and 2 are smaller than those used in the OGE model. Although the
sizes of the wave functions have an influence on the degree of
overlap of the colliding hadrons, the microscopic interactions
$V_C$ and $D_T$ also play a role in the effective meson-baryon
potentials. As we shall discuss shortly ahead, there is an interesting
interplay between these two effects. For a recent discussion on the
influence of hadron sizes on the quark-interchange mechanism, see
Ref~\cite{Vijande}. Regarding the hadron masses, one sees a discrepancy 
of $25$\% (Model~1) and $50$\% (Model~2) on the kaon-nucleon mass 
difference; this does not come as a surprise in view of the 
pseudo-Goldstone boson nature of the kaon, in that a BCS variational 
form for the kaon wave function is a poor substitute for the full Salpeter 
amplitude~\cite{{Bicudo:2001cg},{LlanesEstrada:2001kr}}. On the other
hand, the $DN$ mass difference is within $10$\% of the experimental value 
in both Model~1 and Model~2.

\begin{table}[ht]
\caption{Variational size parameters of the hadron amplitudes and 
hadron mass differences. All values are in MeV.}
\begin{ruledtabular}
\begin{tabular}{c|cccccc} 
 & $\alpha$ & $\beta_K$  & $\beta_D$ & $\Delta M_{NK}$ 
& $\Delta M_{DN}$ & $\Delta M_{DK}$\\
Experiment      &   --     &  --        &  --       & (443) 
& (928) & (1371)   \\   
\hline\\[-.2cm]
Model 1 & 568 & 425 & 508 & 350 & 990 & 1345 \\[0.1cm]
Model 2 & 484 & 364 & 423 & 205 & 1010 & 1220 \\[0.1cm]
OGE~Refs.~\cite{{Hadjimichef:2002xe},{Haidenbauer:2007jq}} & 400 &
350 & 383.5 & -- & -- & --
\end{tabular}
\end{ruledtabular}
\label{tab:sizes}
\end{table}

We are now in position to discuss numerical results for scattering
observables. We solve numerically the Lippmann-Schwinger equation
for the potentials derived from the quark-interchange mechanism
and one-meson exchanges, using the method discussed in Section~2.4
of Ref.~\cite{Adhikari}. The importance of going beyond quark-Born
diagrams by iterating the quark-interchange potentials in a
scattering equation has been stressed
pre\-viously~\cite{{Hadjimichef:2000en},{Hadjimichef:2002xe},
{Haidenbauer:2007jq}}. For the specific case of the OGE
quark-interchange in $\bar DN$ scattering, unitarization of the
scattering amplitude by iteration in a Lippmann-Schwinger equation
leads to a decrease of the cross-section at low energies by a
factor of three as compared to the nonunitarized quark-Born
amplitude~\cite{Haidenbauer:2007jq}.

\vspace{.65cm}
\begin{figure}[htbp]
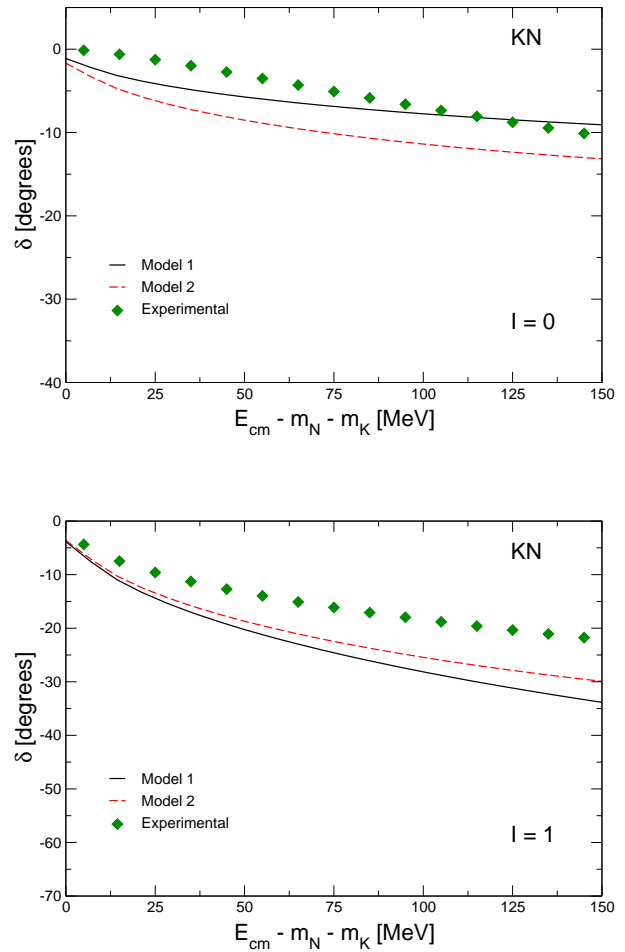

\begin{center}
\resizebox{8.0cm}{!}{\includegraphics{fig3kqm0.eps}}
\\[1.0true cm]
\resizebox{8.0cm}{!}{\includegraphics{fig3kqm1.eps}}
\end{center}
\caption{$KN$ $s$-wave phase shifts for isospin
$I=0$ (upper panel) and $I=1$ (lower panel) channels from the
quark-interchange meson-baryon potential driven by the microscopic
interactions of Model~1 (solid lines) and Model~2 (dashed lines).
The curve with diamond symbols are the results from the phase
shift analysis from the George Washington Data Analysis Center
(DAC), Ref.~\cite{Experdata}.} \label{fig:KN-qm}
\end{figure}

In Fig.~\ref{fig:KN-qm} we present results for the $s$-wave phase 
shifts for the isospin $I=0$ and $I=1$ of the $K^+N$ and $K^0N$ 
reactions. Here we show the results derived from
the baryon-meson potential obtained from the quark-interchange
mechanism driven by the microscopic interactions of Model~1 and
Model~2. Also  shown in the figure are results from phase shift
analysis from the George Washington Data Analysis Center
(DAC)~\cite{Experdata}. As in the case of the model with
OGE~\cite{Hadjimichef:2002xe}, both Model~1 and Model~2 reproduce
the experimental fact that the $s-$wave phase shifts for $I=0$ are
much smaller than for $I=1$. This is due to the combined effects
that for $I=0$, the confining interaction $V_C$ does not
contribute at all and $D_T$ contributes only via diagrams $c$ and
$d$ of Fig.~\ref{fig:qg-inter}.

Also seen in Fig.~\ref{fig:KN-qm}
is the fact that both Model~1 and Model~2 provide large repulsion
-- the confining interaction $V_C$ of Model~2 provides a little
more repulsion than the $V_C$ from Model~1.  As in the case of
OGE, meson-exchanges can be added to obtain a fair description of
the data, as we discuss next.

\begin{figure}[t]
\vspace{.5cm}
\begin{center}
\resizebox{8.0cm}{!}{\includegraphics{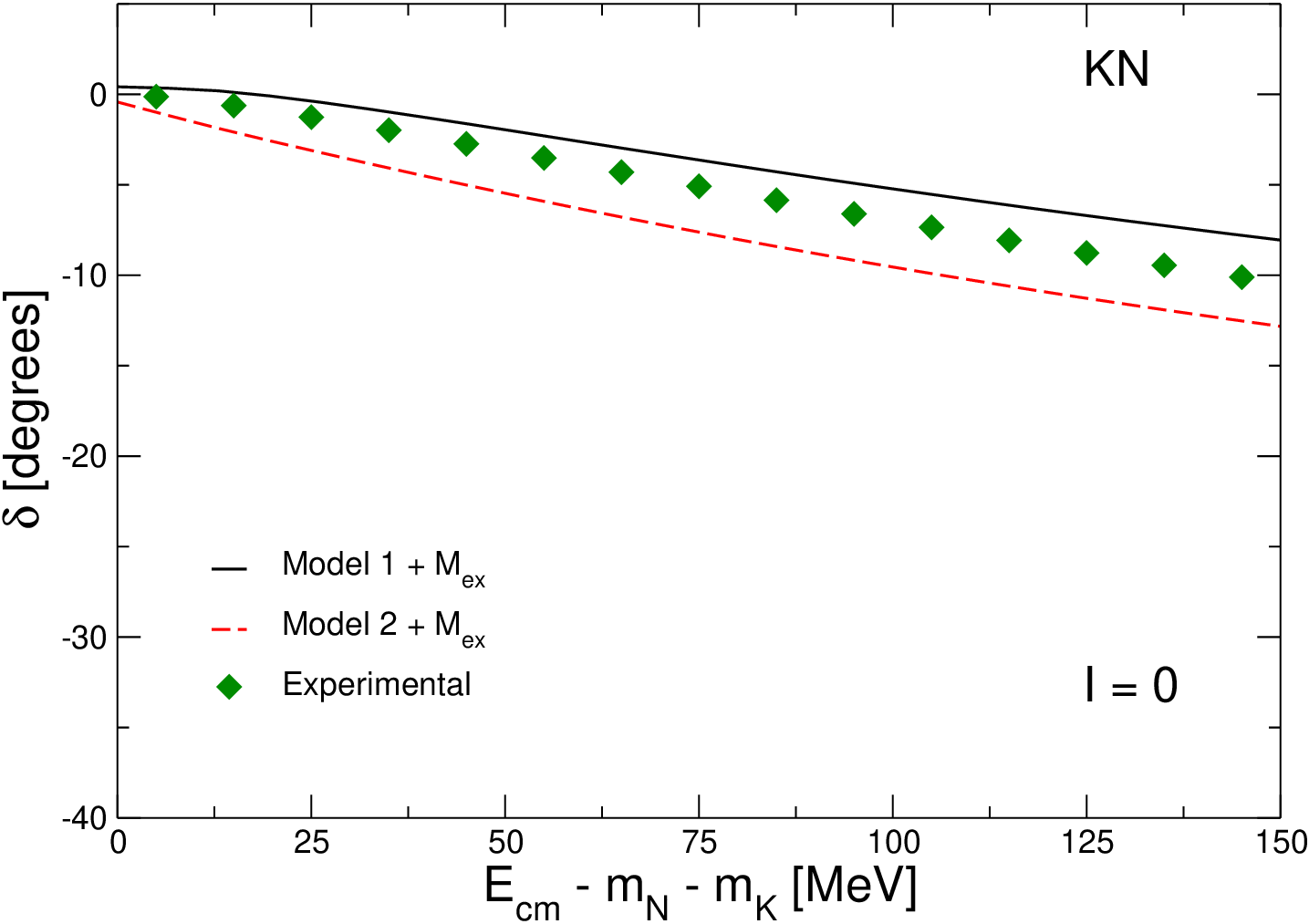}}
\\[1.0true cm]
\resizebox{8.0cm}{!}{\includegraphics{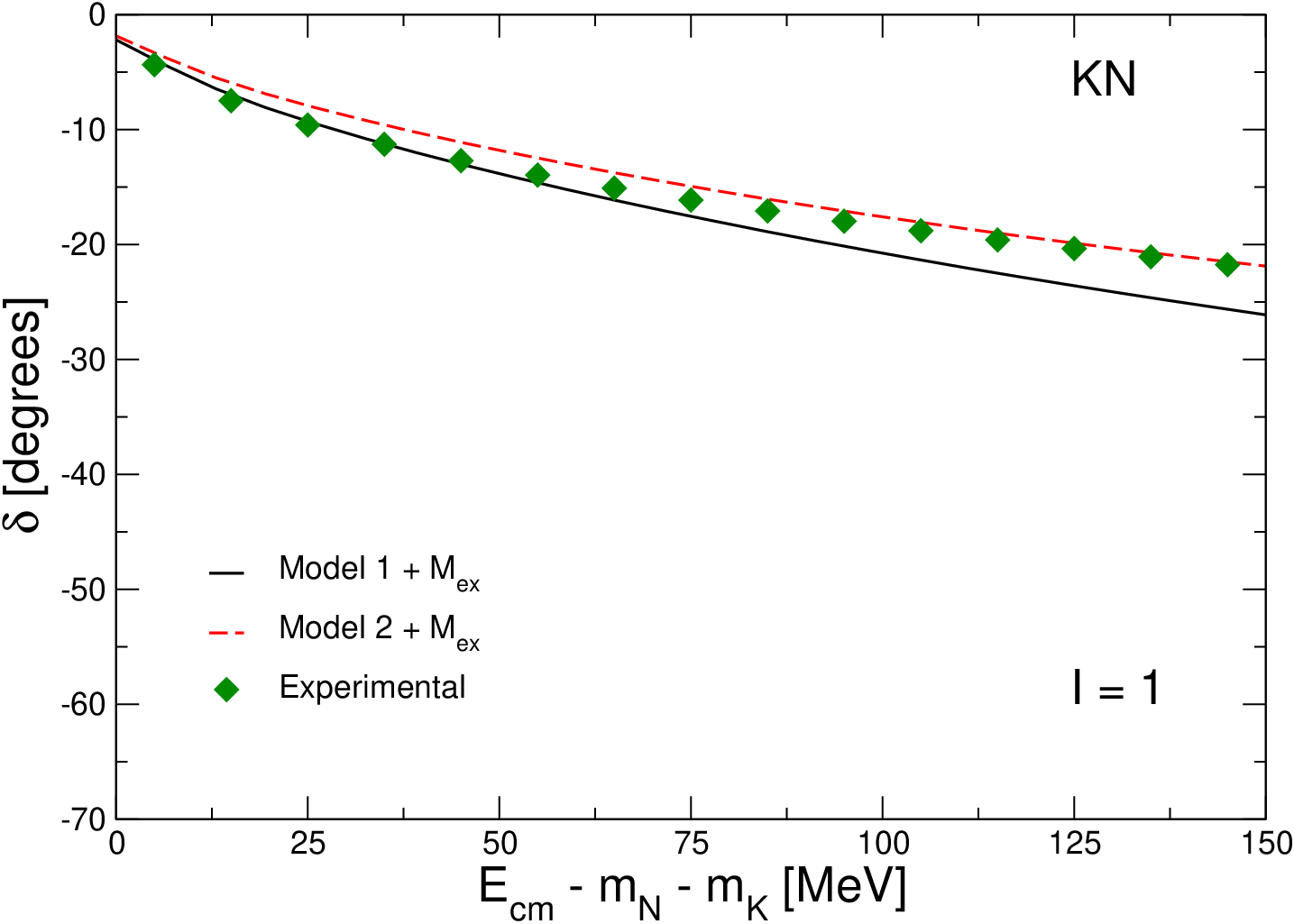}}
\end{center}
\caption{Same as in Fig.~\ref{fig:KN-qm}, but with meson-exchange
potentials added to the quark-interchange meson-baryon potential.}
\label{fig:KN-qm+mex}
\end{figure}

In Fig.~\ref{fig:KN-qm+mex} we present results for the phase
shifts when ($\sigma$, $\omega$, $\rho$, and $a_0$) one-meson
exchanges are added to the quark-interchange potential. The input
parameters for the meson-exchange potential are cutoff masses in
form factors and coupling constants -- we take the values used in
Refs.~\cite{{Haidenbauer:2007jq},{Hadjimichef:2002xe}}, with the
exception of the $\sigma$ and $a_0$ couplings, the product
$g_{\sigma MM} g_{\sigma BB}$ is increased by four and
the product $g_{a_0 MM} g_{a_0 BB}$ is increased by three for a
reasonable description of the phases - this is because the quark-interchange potential
gives strong repulsion. Once the $K^+N$ and $K^0N$ phase shifts
are fitted, we use the same values of the cutoff masses and
couplings to make predictions for the $\bar D N$ system, which we
discuss next.

\begin{figure}[t]
\vspace{.5cm}
\begin{center}
\resizebox{8.0cm}{!}{\includegraphics{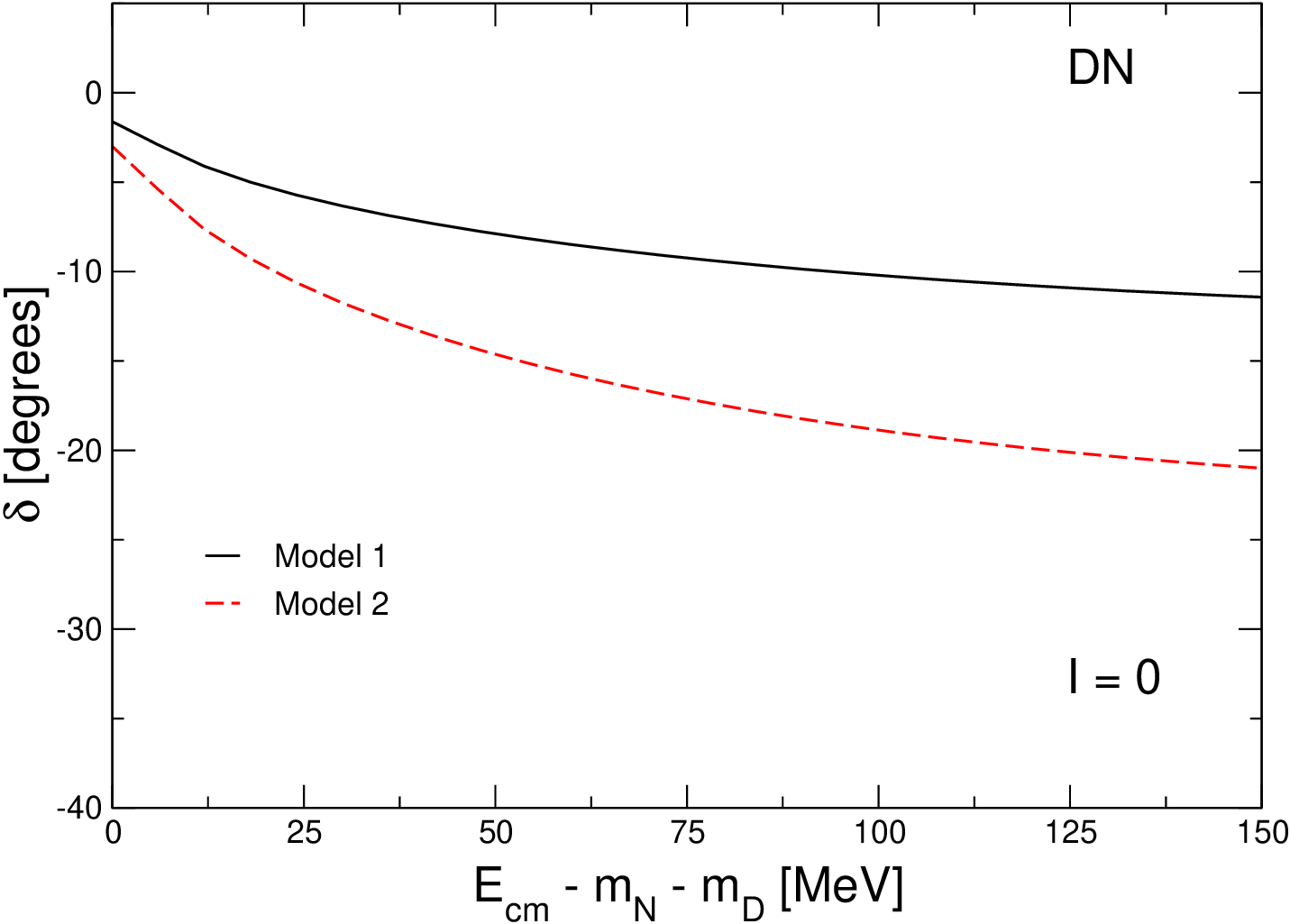}}
\\[1.0true cm]
\resizebox{8.0cm}{!}{\includegraphics{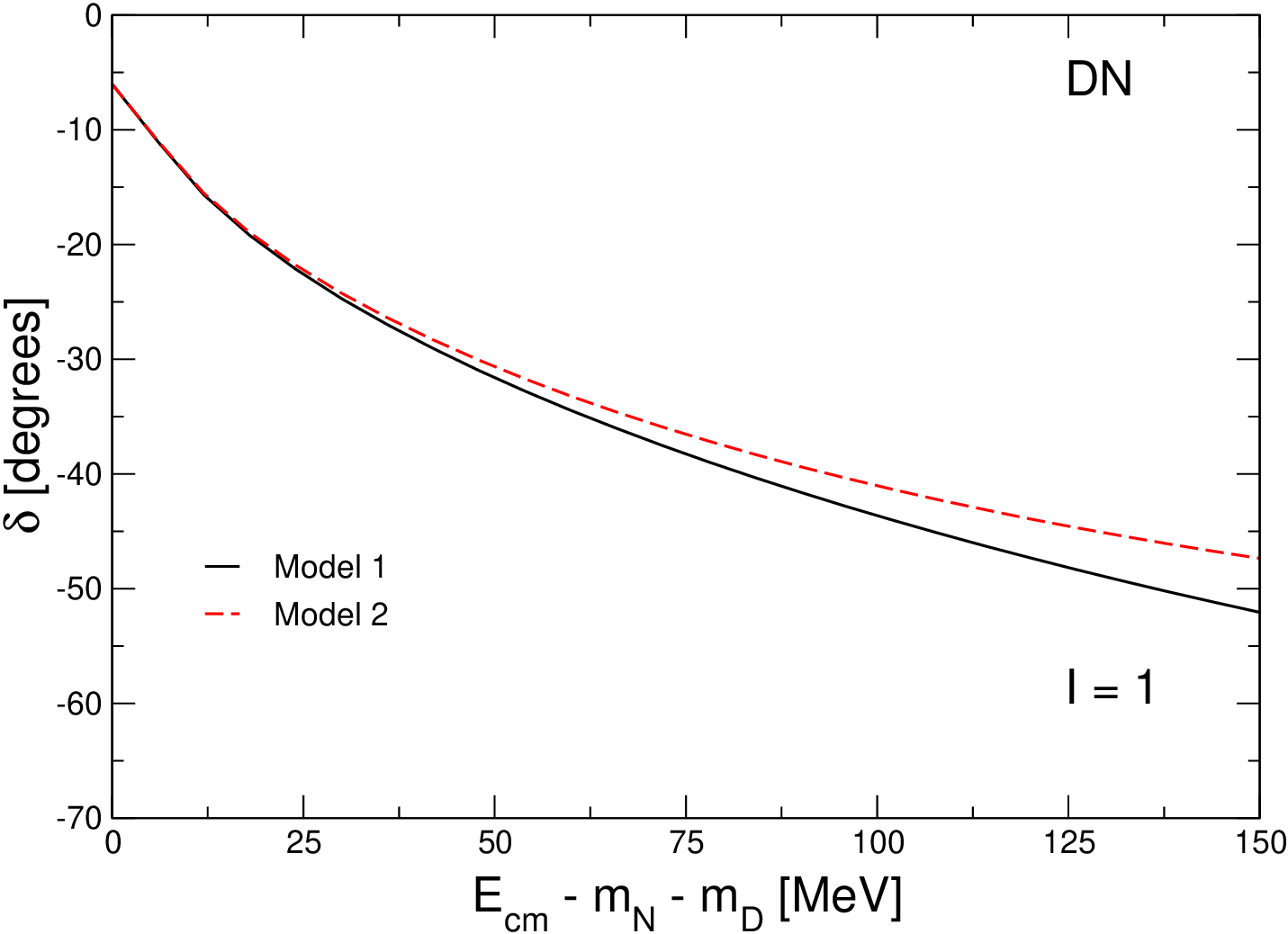}}
\end{center}
\caption{$\bar DN$ $s$-wave phase shifts for $I=0$ (upper panel)
and $I=1$ (lower panel) channels from the quark-interchange
meson-baryon potential driven by the microscopic interactions of
Model~1 (solid lines) and Model~2 (dashed lines).}
\label{fig:DN-qm}
\end{figure}

In Fig.~\ref{fig:DN-qm} we present the $s-$wave phase shifts for
isospin $I~=~0$ and $I=1$ states of the $\bar D^0N$ and $D^-N$
reactions. Results are obtained with a meson-baryon potential from
quark-interchange driven by the interactions of Model~1 and
Model~2. Like in the similar $KN$ system, the phases for the $I=1$
channel are much bigger than those for the $I=0$ channel. Also,
one sees that Model~2 gives a stronger repulsion.

\begin{figure}[t]
\vspace{.5cm}
\begin{center}
\resizebox{8.0cm}{!}{\includegraphics{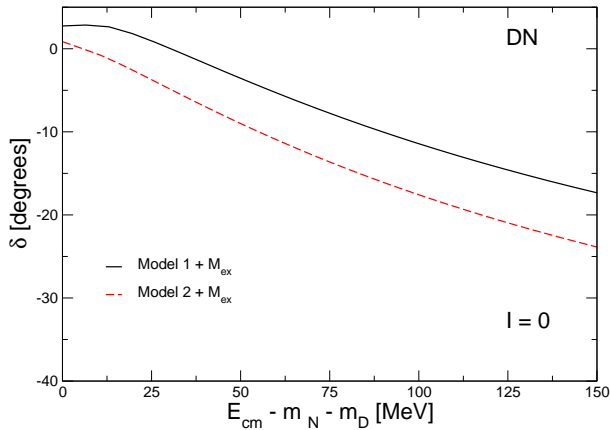}}
\\[1.0true cm]
\resizebox{8.0cm}{!}{\includegraphics{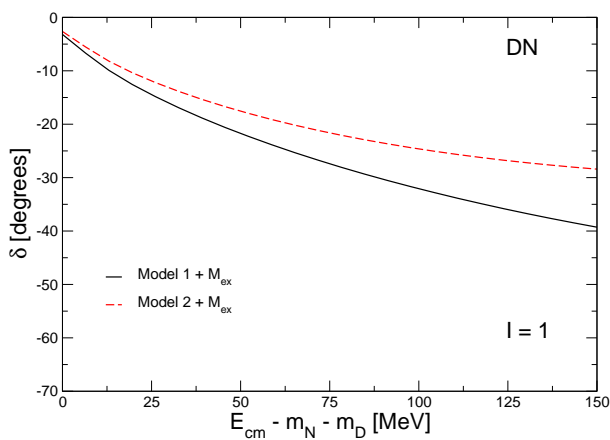}}
\end{center}
\caption{Same as in Fig.~\ref{fig:DN-qm}, but with meson-exchange
potentials added to the quark-interchange meson-baryon potential.}
\label{fig:DN-qm+mex}
\end{figure}

Adding ($\sigma$, $\omega$, $\rho$, and $a_0$) one-meson exchanges
to the quark-interchange potential leads to the results shown in
Fig.~\ref{fig:DN-qm+mex}. Parameters of the meson-exchange
potentials are the same used for the $KN$. The predictions for the
$s-$wave phase shifts for $\bar DN$ system are qualitatively
si\-mi\-lar to the results for the $KN$ system, but are roughly a
factor of two larger than the latter ones.

\begin{figure}[t]
\vspace{.5cm}
\begin{center}
\resizebox{8.0cm}{!}{\includegraphics{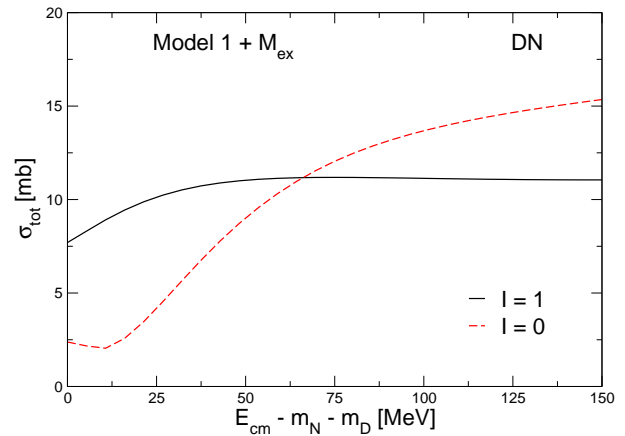}}
\\[1.0true cm]
\resizebox{8.0cm}{!}{\includegraphics{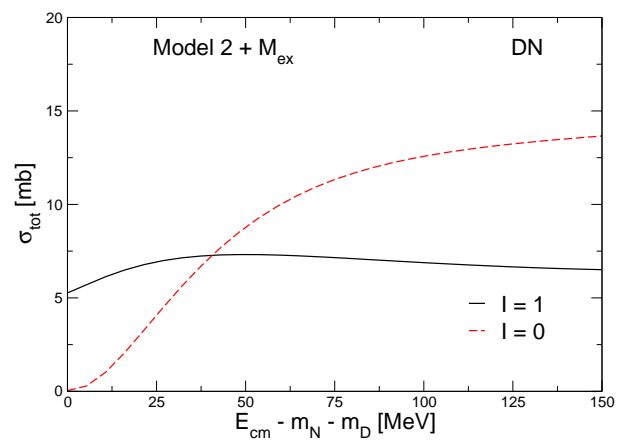}}
\end{center}
\caption{Predictions of $\bar DN$ cross sections for $I=0$ (upper panel)
and $I=1$ (lower panel) using the full meson-baryon potential with
quark-interchange and meson exchanges. }
\label{fig:DN-cross}
\end{figure}

The quark-interchange meson-baryon potential leads to much smaller phase
shifts for the higher partial waves. Like in the $KN$ system, meson exchanges
play a much more important role in these waves. Although not shown here, many
important features of the meson-exchanges discussed in
Ref.~\cite{Haidenbauer:2007jq}, as the interference pattern between
$\rho$ and $\omega$ contributions -- destructive for $I=0$ and constructive
for $I=1$ -- are seen in the present model as well. We do not show these
detailed results here, as they concern the meson exchange part of the full
interaction, but present in Fig.~\ref{fig:DN-cross} the final predictions
for $\bar DN$ cross sections.

The results from both confining models are qualitatively similar, although the
results from Model~1 for the $I=1$ state are on average larger by a factor
roughly equal to two. It is important to note that in the present paper we are
approximating the correlated two-pion exchange contribution by single $\sigma$-meson
exchange. As seen in Ref.~\cite{Haidenbauer:2007jq}, this seems to be a reasonable
approximation for the $I=0$ channel, but underestimates the $I=1$ cross section
by a factor roughly equal to two. We expect a similar feature in the present case.
As a final remark, we note that the results for the cross sections are of the same
order of magnitude as those from OGE quark-interchange, but the shapes of the curves
for the $I=1$ cross sections from Model~1 and Model~2 are di\-fferent at very low
energies from those from OGE. This last feature can possibly be attributed to the
drastically different momentum dependences of $V_C(k)$ and $V_T(k)$ from the
$1/k^2$ dependence of the OGE.

%
\section{Conclusions and Perspectives}
\label{sec:concl}

There is great contemporary interest in studying the interaction of
charmed hadrons with ordinary hadrons and nuclei. Of particular
interest is the study of $D$ mesons in medium, mainly in connection
with the possibility of creating new exotic nuclear bound states by nuclei
capturing charmonia states like $J/\Psi$ and $\eta_c$~\cite{{Brodsky:1989jd},
{Ko:2000jx},{Krein:2010vp},{Tsushima:2011kh}}, or $D$ and $D^*$
mesons~\cite{{Tsushima:1998ru},{GarciaRecio:2010vt},{GarciaRecio:2011xt}}.
One major difficulty here is the complete lack of experimental information
on the free-space interactions that would be of great help to guide model
building -- this is not the case for the si\-mi\-lar problems involving strange
hadrons. In a situation of lack of experimental information, one way
to proceed in model building is to use symmetry constraints,
analogies with other similar processes, and the use of different degrees
of freedom~\cite{{Haidenbauer:2007jq},{Haidenbauer:2008ff},
{Haidenbauer:2009ad},{Haidenbauer:2010ch}}. With such a motivation,
we have implemented a calculation scheme for deriving effective
low-energy hadron-hadron interactions based on a phenomenological
Hamiltonian inspired in the QCD Hamiltonian in Coulomb gauge. The
model Hamiltonian, defined in Eqs.~(\ref{H})-(\ref{Dij}), confines
color and realizes D$\chi$SB. The scheme makes contact with traditional
cons\-tituent quark mo\-dels~\cite{Close:1979bt}, but it goes beyond such
mo\-dels: the constituent quark masses are derived from the very same
interactions that lead to hadron bound states and hadron-hadron interactions,
while in the traditional quark models, quark masses and quark-quark
interactions are specified independently. Moreover, the model confines
color, in that only color-singlet hadron states are of finite energy.
These features of the model are essential for studies seeking signals of
in-medium modifications of hadron properties.

The model Hamiltonian requires as input a Coulomb-like term $V_C$ and
a transverse-gluon interaction $D_T$. We used two analytical forms for
$V_C$: Model~1 is a parametrization of the lattice simulation of
QCD in Coulomb gauge of Ref.~\cite{Voigt:2008rr} and Model~2 was used
in recent studies of glueballs~\cite{Guo:2007sm} and heavy hybrid
quarkonia~\cite{Guo:2008yz}. For $D_T$ we were guided by previous studies
of spin-hyperfine splittings of meson masses~\cite{LlanesEstrada:2004wr}.
Initially we obtained the constituent quark mass function $M_f(k)$ for
different flavors~$f$. Next, we derived an effective low-energy meson-baryon
interaction from a quark-interchange mechanism, whose input is the microscopic
Hamiltonian and hadron bound state wave functions given in terms of quark
degrees of freedom~\cite{{Hadjimichef:1998rx},{Hadjimichef:2000en}}.
We~derived explicit analytical expressions for the effective meson-baryon
interaction by using a low-momentum expansion of the constituent quark mass
function and variational hadron wave functions~\cite{{Bicudo:1991kz},{Bicudo:2001cg}}.
Initially we used the short-ranged quark-interchange potential derived within
the model and added one-meson-exchange potentials to fit experimental
$s-$wave phase shifts of the $K^+N$ and $K^0 N$ reactions. Next, without
changing any parameters besides the strange to charm current quark masses,
$m_s \rightarrow m_c$, we presented the predictions of the model for the
$\bar D^0N$ and $D^- N$ reactions. The results for the cross sections obtained
with the present model are of the same order of magnitude as those from OGE
quark-interchange, of the order of $5$~mb for the $I=1$ state and $10$~mb for
the $I=0$ state, on average. However, the shapes of the curves for the $I=1$
cross sections from Model~1 and Model~2 are dif\-ferent at very low energies from
those from  OGE -- a feature we attribute to the drastically different momentum
dependences  of $V_C(k)$ and $V_T(k)$ from the $1/k^2$ dependence of the OGE.

The model can be improved in several directions. First of all, without sacrificing
the ability of obtaining analytical expressions for the effective meson-baryon
interaction, one can expand the the amplitudes $\psi$ and $\phi$ in a basis of
several Gaussians and diagonalize the resulting Hamiltonian matrix together
with a variational determination of the size parameters of the Gaussians.
Another improvement, that certainly will be necessary for studying in-medium
chiral symmetry restoration, is to use the full mass function $M_f(k)$ for the
light $u$ and $d$ flavors, instead of the low-momentum expansion of
Eq.~(\ref{Mk-app}). This is because at finite baryon density and/or temperature
the mass function $M_f(k)$ for $f=(u,d)$ will loses strength in the
infrared~\cite{{Kocic:1985uq},{Davis:1984xg},{Alkofer:1989vr},{Guo:2009ma},
{Lo:2009ud}} and the expansion in powers of $k^2/M^2$ evidently loses validity.
However, this will have a slight impact on the numerics, since in this case
multidimensional integrals for the determination of the hadron sizes and the
effective meson-baryon interaction need to be performed numerically. The
low-momentum expansion must also be abandoned in the calculation of hadron
wave functions when the quark mass function $M_f(k)$ in the infrared is much
smaller than the average momentum of the quark in the hadron. Such a
situation can happen when $D_T(k)$ is suppressed in the infrared, as in the
study of Ref.~\cite{Pak:2011wu}.

Finally, another subject that needs careful scrutiny is the use of SU(4)
flavor symmetry, Eq.~(\ref{su4}), to fix the coupling constants in the effective
meson Lagrangians. A~recent estimation~\cite{ElBennich:2011py}, using a framework
in which quark propagators and hadron amplitudes are constrained by
Dyson-Schwinger equation studies in QCD, finds that while SU(3)-flavor
symmetry is accurate to 20\%, SU(4) relations underestimate the $DD\rho$
coupling by a factor of $5$. On the other hand, a study employing a
$^3P_0$ pair-creation model with nonrelativistic quark model hadron
wave functions finds smaller SU(4) breakings~\cite{3p0}.

\appendix

\section{Meson-exchange contributions}
\label{app:mex}

The meson-exchange contributions are obtained from the following effective
Lagrangian densities~\cite{{Lin:1999ad},{Lin:2000jp}}:
%
\begin{eqnarray}
\hspace*{-1.cm}\mathcal{L}_{NNS}(x) = g_{NNS} \bar{\psi}^{(N)}(x)
{\bm\tau}\cdot{\bm\phi}^{(S)}(x) \psi^{(N)}(x) ,
\label{a0-sigma}
\end{eqnarray}
%
\begin{eqnarray}
\hspace*{-.4cm}\mathcal{L}_{NNV}(x) &=& g_{NNV}
\Bigl[ \bar{\psi}^{(N)}(x) \gamma^{\mu} \psi^{(N)}(x)
\bm{\tau}\cdot\bm{\phi}_{\mu}^{(V)}(x)
\nonumber\\
&&\hspace*{-1.75cm} + \,\left(\frac{\kappa_{V}}{2M_{N}}\right)
\bar{\psi}^{(N)}(x) \sigma^{\mu\nu} \psi^{(N)}(x)
\bm{\tau}\cdot\partial_{\nu}\bm{\phi}_{\mu}^{(V)}(x) \Bigr],
\label{NN-rho-omega}
\end{eqnarray}
%
\begin{eqnarray}
\hspace*{-1.6cm}\mathcal{L}_{PPS}(x) = g_{PPS} \varphi^{(P)}(x) \phi^{(S)}(x)
\varphi^{(P)}(x) ,
\label{DDsigma}
\end{eqnarray}
%
\begin{eqnarray}
\hspace*{-2.cm}\mathcal{L}_{PPV}(x) = ig_{PPV} \,
\Bigl[\varphi^{(P)}(x) \,
\left(\partial_{\nu}\varphi^{(P)}(x)\right)\bm{\tau}
\nonumber\\
- \, \left(\partial_{\nu}\varphi^{(\bar{P})}(x)\right) \,
\bm{\tau}\varphi^{(P)}(x)
\Bigr]\cdot{\bm{\phi}^{(V)}_{\nu}}(x) .
\label{rho-omega}
\end{eqnarray}
In these, $\psi^{(N)}$ denotes the nucleon doublet, $\phi^{(P)}(x)$ the
charmed and strange meson doublet, $\bm{\phi}_{\mu}^{(V)}(x)$ the iso-triplet of
$\rho$ mesons, and $\bm{\tau}$ are the Pauli matrices. The Lagrangians
for the $\sigma$ and $\omega$ mesons are obtained taking $\bm{\tau}\rightarrow 1$
in the expressions above and in addition $\kappa_{V}=0$ for the case of $\omega$.

The tree-level potentials derived from the above Lagrangian densities lead to
the following expressions for the vector-meson exchanges ($v=\rho,\omega$):
\begin{eqnarray}
V^{v}(\p,\p') &=& \frac{g_{NNv} \, g_{PPv}}
{(2\pi)^{3}\sqrt{4\omega(p) \omega(p')}} \, (p'+p)_{\mu} \, \Delta^{\mu\nu}_v(q)
\nonumber\\
&&\hspace*{-1.0cm}\times \, \left[A_\nu(ps,p's')
+\left(\frac{\kappa_v}{2m_N}\right)B_\nu(ps,p's')\right],
\label{Vv}
\end{eqnarray}
and for the scalar-meson exchanges ($S=\sigma,a_0$)
\begin{equation}
V^{S}(\p',\p) = \frac{g_{NNS} \,
g_{PPS}} {(2\pi)^{3}\sqrt{4 \omega(p) \omega(p')}}\, \Delta_{S}(q)
\bar{u}(\p',s')u(\p,s),
\label{Vs}
\end{equation}
where $m_N$ is the nucleon mass and $\omega (q) = (q^{2} + m^2)^{1/2}$ with
$m$ the meson masses, $\Delta^{\mu\nu}_v(q)$ and $\Delta_S(q)$ are the
vector-meson and scalar-meson propagators, $u(\p, s)$ are the Dirac spinors
of nucleons (same expression as in Eq.~(\ref{u}), with $M(k)$ replaced by
the nucleon mass $m_N$), and the quantities $A_{\mu}$ and $B_{\mu}$ are given
by
\begin{eqnarray}
A_\mu(ps,p's') &=& \bar{u}(\p',s')\gamma_{\mu}u(\p,s),
\label{Amu} \\[0.3cm]
B_{\mu}(ps,p's') &=& \bar{u}(\p',s')\,i\sigma_{\mu\nu}\,
q^{\nu}\,u(\p,s).
\label{Bmu}
\end{eqnarray}
To avoid divergences in the Lippmann-Schwinger equation, the meson-exchange
potentials are regularized phenomenologically by monopole form factors at
each vertex:
\begin{equation}
F_{i}(\bm{q}^{2})=\left(\frac{\Lambda_{i}^{2}-m_{i}^{2}}
{\Lambda_{i}^{2}+\bm{q}^{2}}\right)
\end{equation}
where $\q =\p'-\p$, $m_{i}$ is the mass of the exchanged meson and
$\Lambda_{i}$ is a cutoff mass. The coupling constants are fixed by SU(4)
symmetry as in Ref.~\cite{Haidenbauer:2007jq}:
\begin{equation}
g_{\bar{D}\bar{D}\rho} = g_{\bar{D}\bar{D}\omega} =g_{KK\rho}= g_{KK\omega}=
\frac{g_{\pi\pi\rho}}{2}.
\label{su4}
\end{equation}
%


\end{document}